\documentclass[useAMS,usenatbib]{mn2e}
\usepackage{amsmath}
\usepackage{graphicx}
\usepackage{xcolor}
\numberwithin{equation}{section}

\newcommand{\be}{\begin{equation}}
\newcommand{\ee}{\end{equation}}

\newcommand{\e}{\times10^}
\newcommand{\Mmin}{M_{\rm{CO,min}}}

\newcommand{\nuco}{\nu_{\rm{CO}}}

\newcommand{\co}{\rm{CO}}

\newcommand{\tco}{\left<T_{\co}\right>}
\newcommand{\nuobs}{\nu_{\mathrm{obs}}}
\newcommand{\zco}{z_{\mathrm{CO}}}
\newcommand{\fduty}{f_{\mathrm{duty}}}
\newcommand{\sfr}{\mathrm{SFR}}

\begin{document}

\title[Carbon monoxide intensity mapping at moderate redshifts]{Carbon monoxide intensity mapping at  moderate redshifts}
\author[Patrick C. Breysse, Ely D. Kovetz, and Marc Kamionkowski]{Patrick C. Breysse,$^{1}$\thanks{pbreysse@pha.jhu.edu (PCB); kamion@pha.jhu.edu (MK)} Ely D. Kovetz,$^{2}$ \thanks{elykovetz@gmail.com (EDK)} and Marc Kamionkowski$^{1}$\footnotemark[1] \\
$^{1}$ Department of Physics and Astronomy, Johns Hopkins University, Baltimore, MD 21218 USA \\
$^{2}$ Theory Group, Department of Physics and Texas Cosmology Center, The University of Texas at Austin, Texas 78712, USA}

\maketitle
\label{firstpage}

\begin{abstract}
We present a study of the feasibility of an intensity-mapping survey targeting the 115 GHz CO(1-0) rotational transition at $z\sim3$.  We consider four possible models and estimate the spatial and angular power spectra of CO fluctuations predicted by each of them.  The frequency bandwidths of most proposed CO intensity mapping spectrographs are too small to use the Limber approximation to calculate the angular power spectrum, so we present an alternative method for calculating the angular power spectrum.  The models we consider span two orders of magnitude in signal amplitude, so there is a significant amount of uncertainty in the theoretical predictions of this signal.  We then consider a parameterized set of hypothetical spectrographs designed to measure this power spectrum and predict the signal-to-noise ratios expected under these models.  With the spectrographs we consider we find that three of the four models give an SNR greater than 10 within one year of observation.  We also study the effects on SNR of varying the parameters of the survey in order to demonstrate the importance of carefully considering survey parameters when planning such an experiment. 
\end{abstract}

\begin{keywords}
cosmology: theory -- cosmology: large-scale structure of universe -- cosmology: diffuse radiation
\end{keywords}

\section{Introduction}
Over the last decade studies of the cosmic microwave background from experiments like WMAP \citep{wmap9} and Planck \citep{planck} have provided unparalleled insight into the structure of the universe at the surface of last scattering.  Meanwhile, large galaxy surveys such as the Sloan Digital Sky Survey \citep{sdss} have probed the structure of the universe at low redshifts.  However, there is a large period of cosmic history for which we have very little information from direct observations.  At some point, galaxies become too faint to detect individually.  A relatively new technique known as intensity mapping has arisen as a way to fill this gap.  First proposed by \citet{sss}, intensity mapping involves studying the large scale fluctuations in the intensity of a given spectral line emitted by a large number of unresolved objects.  Since more distant emitters will be more highly redshifted, this technique could allow the study of the three dimensional structure of the universe.

Intensity mapping can be performed using many different spectral lines.  The most commonly discussed is the 21 cm neutral hydrogen line \citep{fob}.  However, other lines trace different physical processes, and some lines may be easier to study, either because they are brighter or they appear in a less difficult frequency band.  Thus it is useful to study other lines in addition to the 21 cm line.  Some other lines which have been proposed include CII \citep{gcs}  and Ly$\alpha$ \citep{pullenb}.  Here we focus on the rotational transitions of carbon monoxide, particularly the lowest order transition CO(1-0) at 115 GHz.  CO forms primarily in star forming regions, so intensity mapping with this line provides information on the spatial distribution of star formation in the universe.  In addition, the frequencies of the CO transitions fall within a range where existing infrastructure could be adapted to studying it.  CO intensity fluctuations were first studied by \citet{righi} as a foreground contaminant to CMB measurements, then later as a tracer of large scale structure \citep{vl,lidz,pullen}.

In this paper we expand on the work of \citet{pullen} and study the feasibility of a CO intensity mapping   survey targeted at $z\sim3$.  Since this is the redshift where the cosmic star formation rate is expected to be highest \citep{hb}, this is a good place to attempt a first detection of this signal.  Once the techniques for measuring and analyzing this signal are demonstrated successfully at this moderate redshift, surveys could be performed at redshifts corresponding to other periods, such as the epoch of reionization \citep{lidz,mf}.  We consider four models for cosmological CO emission from the literature and estimate the strength of the CO signal predicted by each.  In doing so, we demonstrate a simple one-parameter family of models which can account for a wide variety of assumptions about CO emission.  When we calculate the power spectra using these models, we find that the difference in amplitude between the largest amplitude signal and the smallest is roughly two orders of magnitude, which clearly illustrates the lack of theoretical understanding of star formation physics at high redshifts.  When studying the prospects for an experiment to detect this signal, we explore the effect of the instrumental parameters on the predicted signal-to-noise ratio and demonstrate the importance of carefully considering the values of these parameters when designing a survey.  In particular, attempting to survey too large of an area of the sky or choosing a spectrograph with insufficient resolution can significantly decrease the chance of detecting the CO signal.

In Section 2 below we will summarize four models from the literature and use them to estimate the power spectrum of CO fluctuations.  In Section 3 we will estimate the signal-to-noise ratios that would be obtained with an optimal survey under the assumptions of these models.  We conclude in section 4.  For all of the calculations presented below we use the \citet{tinker} mass function and the following $\Lambda$CDM cosmological parameters: $\left(\Omega_b,\ \Omega_m,\ \Omega_\Lambda,\ h,\ \sigma_8,\ n_s\right)=\left(0.046,\ 0.27,\ 0.73,\ 0.7,\ 0.8,\ 1\right)$.

\section{Modeling CO Emission}
In this section we will outline our method for modeling the power spectrum of CO fluctuations.

\subsection{Deriving the CO Temperature}
When calculating the average CO brightness temperature, we follow the method presented by \citet{lidz}, which is summarized here.  The specific intensity of a CO line observed at frequency $\nuobs$ at $z$=0 can be found from the radiative-transfer equation,
\be
\label{si}
I(\nuobs)=\frac{c}{4\pi}\int_0^\infty\epsilon\left[\nuobs(1+z')\right]\frac{dz'}{H(z')(1+z')^4},
\ee
where $H(z)$ is the Hubble parameter and $\epsilon\left[\nuobs(1+z')\right]$ is the proper volume emissivity of the line in question.  CO is emitted from within halos, so it is natural to calculate the volume emissivity from a halo luminosity function.  We use a simple estimate of the specific luminosity,
\be
\label{lfun}
L_\mathrm{CO}=A\delta(\nu-\nuco)\left(\frac{M}{M_{\sun}}\right)\ L_{\sun},
\ee
which is linear in halo mass.  We have assumed that the targeted line is a Dirac delta function emitted at frequency $\nuco$.  The parameter $A$ is an overall normalization which we will calculate in section 2.1.  We define a minimum mass $\Mmin$ below which halos are too small to emit CO lines, and we assume that a fraction $\fduty$ of halos more massive than $\Mmin$ are emitting at any given time.  For a given halo mass function $dn/dM$ the volume emissivity is then
\begin{multline}
\epsilon(\nu,z)=A\delta(\nu-\nuco)(1+z)^3\fduty \\ \times\int_{\Mmin}^\infty M\frac{dn(z)}{dM}dM.
\end{multline}
For CO lines emitted at redshift $\zco$, this gives a specific intensity
\be
I(\nuobs)=\frac{A}{4\pi}\frac{1}{\nuco}\frac{c}{H(\zco)}\fduty\int_{\Mmin}^\infty M\frac{dn(\zco)}{dM}dM,
\ee
or, written as a brightness temperature,
\begin{multline}
\label{intensity}
\tco=\frac{A}{8\pi}\frac{1}{\nuco^3}\frac{c^3}{2k_BH(\zco)}\fduty(1+\zco)^2 \\ \times\int_{\Mmin}^\infty M\frac{dn(\zco)}{dM}dM.
\end{multline}

\subsection{Theoretical Models}
With the above expression for the average CO temperature, we now need a model of CO emission which can allow us to calculate the parameter $A$.  In this section, we consider four such models.  The first and simplest is proposed by \citet{vl}, hereafter referred to as VL10.  They estimate the star-formation rate for a halo of mass $M$ by assuming that a fraction $f_\star=0.1$ of the baryons in a halo form stars at a constant rate over a time period $t_s\approx10^8$ years, where $\fduty$ is the ratio of this time to the Hubble time at redshift $\zco$.  This gives a star formation rate (SFR) of
\be
\sfr=\frac{f_\star}{t_s}\frac{\Omega_b}{\Omega_m}M.
\ee
With this relation, we can determine the CO luminosity of a halo if we have a relationship between SFR and $L_{\mathrm{CO}}$.  In this model, this relation is obtained by measuring the ratio of CO luminosity to SFR from M82, which is observed to be $3.7\e3\ L_{\sun}/(M_{\sun}/\rm{yr})$ \citep{m82}.  
Combining these scaling relations allows us to set the value of the parameter $A$ from equation (\ref{lfun}).  The value in this model is
\be
A_{\rm{VL10}}=6.24\e{-7}.
\ee
This value can be used with equation (\ref{intensity}) to determine the average brightness temperature.

The second model we consider is Model A from P13.  The CO luminosity function is calculated similarly to the VL10 model, but instead of using the M82 normalization, they use a set of empirical scaling relations described by \citet{car}.  They first relate CO luminosity to FIR luminosity, then FIR luminosity to SFR, and then SFR to halo mass in a similar manner to VL10.  This gives a luminosity function which is still linear in mass, but with a different normalization
\be
\label{P13A}
A_{\rm{P13A}}=2\e{-6}.
\ee

Model B from P13 uses a slightly different method to calculate the CO brightness temperature.  Instead of trying to calculate a luminosity function, this model assumes that the star formation rate $S$ follows the Schechter function:
\be
\Phi(S)dS = \phi_\star\left(\frac{S}{S_\star}\right)\exp\left(-\frac{S}{S_\star}\right)\frac{dS}{S_\star},
\ee
where $\phi_\star$ is a characteristic density and $S_\star$ is a characteristic star formation rate \citep{sch}.  Integrating this function gives the cosmic SFR density, which can then be combined with the same SFR-CO luminosity scaling relation used in model P13A to get an estimate of the CO volume emissivity.  This emissivity can then be entered into equation (\ref{si}) to get the CO brightness temperature.  Though the calculation of $\tco$ in this model is somewhat more involved than the one described in equations (\ref{si})-(\ref{intensity}), we can get a reasonable estimate of the brightness temperature in model P13B simply by adjusting the value of $A$.  P13 state that $\tco$ in model B is roughly a factor of 4.8 higher than in model A at $z\sim3$, so the brightness temperature in model P13B can be calculated using equations (\ref{si})-(\ref{intensity}) with
\be
A_{\rm{P13B}}=9.6\e{-6}.
\ee

The final model we look at here is proposed by \citet{righi}, which we will refer to as R08.  Instead of just assuming that some fraction of halos are forming stars at any given time, the R08 model assumes that star forming episodes happen following major merger events.  They estimate that the mass $M_\star$ of stars formed when two halos of mass $M_1$ and $M_2$ merge into a halo of mass $M$ is
\be
M_\star=4\frac{\Omega_b}{\Omega_m}f_\star\frac{M_1M_2}{M}.
\ee
From there, one can calculate the merger rates for halos of a given mass and integrate over all possible masses to determine the total star formation rate.  The authors then use the same M82 normalization from the VL10 model to calculate the CO luminosity.  As with model P13B, the full calculation of $\tco$ in R08 is more complicated than what we have shown here thus far, but we can obtain a good approximation of their result by choosing the correct value for $A$.  In this case the necessary value is approximately twice the one used for model P13A
\be
A_{\rm{R08}}=4\e{-6}.
\ee

The expression for CO brightness temperature from equation (\ref{intensity}) can be rewritten as
\begin{multline}
\tco(z)=0.60\left[\frac{A}{2\e{-6}}\right]\left[\frac{2.2\e9\ \mathrm{yr}}{t_H}\right]\left[\frac{H(z=3)}{H(z)}\right] \\ \times\left[\frac{1+z}{4}\right]^2\left[\frac{\int_{\Mmin}^\infty M\frac{dn}{dM}dM}{7.05\e9\ M_{\sun}/\mathrm{Mpc}^3}\right]\ \mathrm{\mu K},
\end{multline}
where the numerical values are given for model P13A targeted at $z=3$ a minimum halo mass of $\Mmin=10^9\ M_{\sun}$.  Table \ref{tcotable} gives the fiducial values of $\tco$ for the four models above.  It is clear that there is a large amount of theoretical uncertainty regarding the amplitude of the expected CO signal.  This justifies our decision to simplify our estimation of $\tco$ in models P13B and R08, since the differences between our calculations and the full calculations will be considerably smaller than the differences between the models.  For the remainder of this paper we will consider only model P13A at $z=3$ unless stated otherwise.  However, the reader should bear in mind the broad range of theoretical possibilities when following the rest of our results.

\begin{table}
\begin{center}
\caption{Fiducial $\tco$ values at redshift 3 for each of the four models we consider.}
\label{tcotable}
\begin{tabular}{c || c}
\hline
Model & $\tco$ ($\mu$K) \\
\hline
VL10 & 0.19 \\
P13A & 0.60 \\
R08 & 1.20 \\
P13B & 2.88 \\
\hline
\end{tabular}
\end{center}
\end{table}

\subsection{The CO Power Spectrum}
A simple way to estimate the power spectrum of CO emission is given in \citet{lidz}.  For a given $\tco$, the three dimensional power spectrum should take the form
\be\label{eq:PowerSpectrum}
P_{\co}(k,z)=\tco^2(z) \left[b^2(z)P_m(k,z)+P_{\rm{shot}}(z)\right].
\ee
The first term gives the contribution of the power spectrum from the clustering of matter in the Universe.  Since CO is emitted from within halos, this term is simply the halo power spectrum scaled by $\tco$.  The halo power spectrum is calculated by multiplying the linear power spectrum $P_m(k,z)$ by a mass-averaged bias factor
\be
b(z)=\frac{\int_{\Mmin}^\infty M\frac{dn}{dM} b(M,z)dM}{\int_{\Mmin}^\infty M\frac{dn}{dM}dM}.
\ee
For the mass-dependent bias, $b(M,z)$, we use
\be
b(M,z)=1+\frac{\nu(M,z)-1}{\delta_c},
\ee
where $\delta_c=1.69$, $\nu(M,z)=\delta_c/\sigma(M,z)$, and $\sigma(M,z)$ is the RMS density fluctuation in a spherical region containing mass $M$ \citep{mw}.  The second term of equation (\ref{eq:PowerSpectrum}) is the shot noise contribution from the random distribution of halos on the sky. This contribution can be expressed as \citep{lidz}
\be
P_{\rm{shot}}(z)=\frac{1}{f_{\rm{duty}}}\frac{\left<M^2\right>}{\left<M\right>^2},
\ee
where
\be
\left<M^2\right>=\int_{\Mmin}^\infty M^2\frac{dn}{dM}dM,
\ee
and
\be
\left<M\right>=\int_{\Mmin}^\infty M\frac{dn}{dM}dM.
\ee
Note that the shot noise has no dependence on wavenumber.

The exact expression for converting a spatial power spectrum $P(k)$ to an angular spectrum $C_\ell$ is
\be\label{eq:angularPS}
C_\ell=\frac{2}{\pi}\int k^2P(k)\left[\int f(r)j_\ell(kr)dr\right]^2dk,
\ee
where $r$ is the comoving distance, $j_\ell(kr)$ is the spherical bessel function and $f(r)$ is the selection function which is determined by the frequency bandwidth of the instrument used to observe the CO emission.  For simplicity, we use a top hat $f(r)$ for this analysis.  

Equation (\ref{eq:angularPS}) is somewhat time consuming to evaluate numerically, so we take steps to simplify the calculation somewhat.  One method commonly referred to as the Limber approximation \citep{limber,rubin} is obtained by assuming the power spectrum is a slowly varying function of $k$
\be
\label{cllimber}
C_\ell=\int \frac{H(z)}{c}\frac{f^2(z)}{r^2(z)}P_{\co}[k=\ell/r(z),z]dz.
\ee
However, this approximation is only valid when the width $\delta r$ of the selection function satisfies $\ell\delta r/r\gg 1$, i.e. when the width of the observed shell is large compared to the scale of fluctuations being considered.  For small $\ell$'s and small frequency bandwidths the Limber approximation fails.  In the regime where $\ell\delta r/r\ll 1$, we can instead simplify the calculation by approximating $f(r)$ as a delta function.  The angular power spectrum for a selection function centered on redshift $z_0$ is then given by
\be
\label{cllow}
C_\ell=\frac{2}{\pi}\int k^2P(k)j_\ell^2[kr(z_0)]dk.
\ee
Some attempts have been made to simplify this expression further by assuming $P(k)$ is a power law \citep{zfh}, but that approximation only works if the slope of the power law is less than 2.  Since the slope of $P_{\co}(k)$ is near 3 in the range of interest, we cannot use this approximation.

\begin{figure}
\centering
\includegraphics[width=\columnwidth]{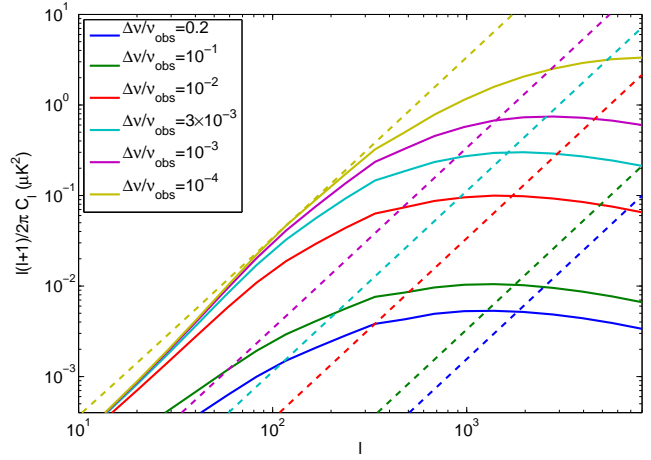}
\caption{Clustering term of the angular CO power spectrum at $z_0=3$ for instruments with different frequency bandwidths.  The solid lines show the contribution from the clustering term and the dashed lines show the contributions from shot noise.  The clustering term loses its dependence on frequency bandwidth when the width of the spatial shell being observed becomes smaller than the size of the features being probed at a given $\ell$ value.}
\label{Cl}
\end{figure}

\begin{figure}
\centering
\includegraphics[width=\columnwidth]{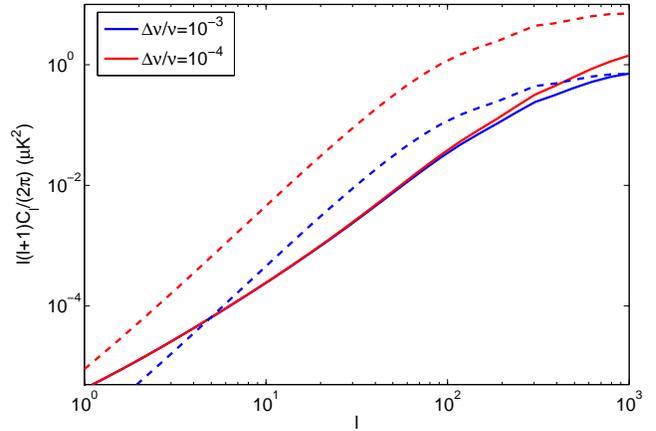}
\caption{Comparison of full power spectrum (solid) with Limber approximation (dashed) for two narrow frequency bandwidths.  The Limber approximation fails in this case because the width of the redshift shell being probed is small compared to the spatial size of the brightness fluctuations.}
\label{limbercompare}
\end{figure}

We calculate the full angular power spectrum by using equations (\ref{cllimber}) and (\ref{cllow}) in the areas where they are valid and interpolating between them.  Figure~\ref{Cl} shows the clustering (solid lines) and shot noise (dashed lines) terms of the angular power spectrum for our fiducial model P13A for instruments with different bandwidths. These power spectra are calculated at $z=3$.  As the bandwidth increases, the amplitude of the signal falls off sharply.  This makes it very difficult to find this signal in WMAP or Planck data since those instruments had very wide frequency bands.  This is likely why the attempt in P13 to find the CO signal in WMAP data was unsuccessful.  Also note that at small bandwidths, the clustering term approaches a maximum amplitude.  This is due to the use of equation (\ref{cllow}) at low $\ell$'s.  If only the Limber approximation was used, the clustering term would continue to increase beyond the limit seen here.  Figure \ref{limbercompare} shows this effect for the two narrowest bandwidths from Figure \ref{Cl}.

\section{Signal-to-Noise Estimation}
We now introduce a sample instrument to measure the CO signal and estimate the signal-to-noise ratio it would produce.  We consider a spectrograph with a 1 GHz total bandwidth targeted at $z=3$. This bandwidth is split into 35 channels, giving a spectral resolution $R=1000$.  We assume an observation time $t_{\rm{obs}}=1$ yr and choose the detector sensitivity $s$ and beam size $\theta_{\rm{fwhm}}$ to be $800\ \mu\rm{K}\sqrt{\rm{s}}$ and 10 arcminutes respectively.  These values are comparable to those for experiments currently under consideration (see, for example, \citet{vtl} and P13).  For a survey covering a solid angle $\Omega_s$, the instrumental noise can be modeled as a random field on the sky with a power spectrum
\be
\label{cln}
C_\ell^n=\frac{s^2\Omega_s}{t_{\rm{obs}}B_\ell^2}
\ee
\citep{tegmark}, where $B_\ell$ is the beam profile, typically approximated as a Gaussian,
\be
B_\ell=e^{-\theta_{\rm{fwhm}}^2\ell(\ell+1)/(16\ln2)}.
\ee

The angular power spectrum can be measured in each channel of this spectrograph, after which the signals can be stacked to increase the overall SNR.  The SNR for such an instrument with $N_{ch}$ channels is given by
\be
\label{snr}
\textrm{SNR}^2 = N_{ch}\sum_\ell\frac{C_l^2}{\sigma_\ell^2}.
\ee
where
\be
\sigma_{\ell}=\sqrt{\frac{8\pi}{\Omega_s(2\ell+1)}}C_\ell^n,
\label{varnull}
\ee
\citep{jkw}.  Combining equations (\ref{snr}) and (\ref{varnull}) with equation (\ref{cln}) yields
\begin{multline}
\label{snrnull}
SNR^2=\frac{N_{ch}}{8\pi}\left(\frac{A}{A_{\rm{P13}}}\right)^4\frac{t^2_{\rm{obs}}}{s^4\Omega_s}\sum_\ell(2\ell+1) \left(C_\ell^{\rm{P13A}}\right)^2 \\ \times e^{\theta^2_{\rm{fwhm}}\ell(\ell+1)/(4\ln2)},
\end{multline}
where $C_\ell^{\rm{P13A}}$ is the angular power spectrum in the P13A model.  

This expression is only valid if one is testing against the null hypothesis, i.e. if one is only interested in seeing if this signal exists at all.  It is useful for a first detection attempt, but when trying to obtain useful cosmological information from a signal it is necessary to include an extra cosmic variance term in equation (\ref{varnull}) to account for the limited number of modes available in the survey.  The signal variance in this case is given by
\be
\sigma_{\ell}=\sqrt{\frac{8\pi}{\Omega_s(2\ell+1)}}\left(C_\ell+C_\ell^n\right).
\label{varcv}
\ee

\begin{figure}
\centering
\includegraphics[width=\columnwidth]{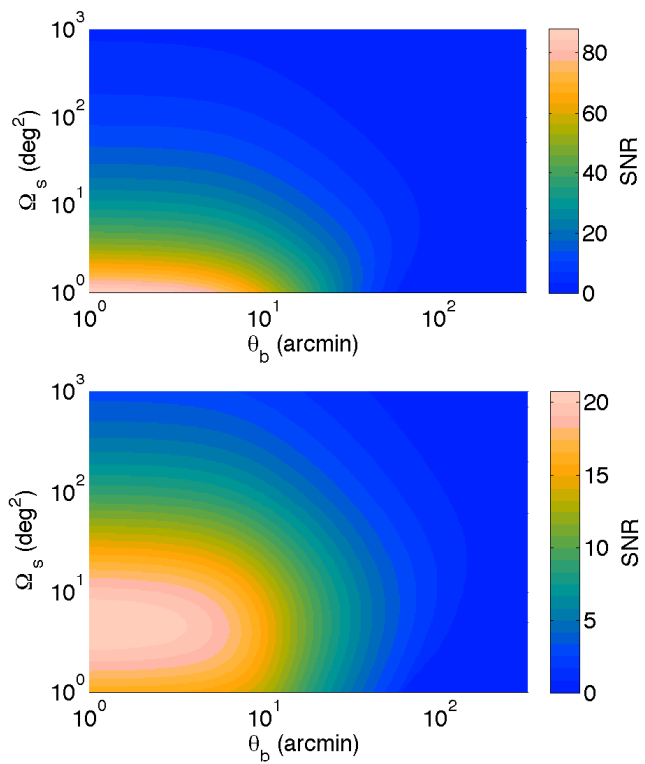}
\caption{SNR as a function of beam size and survey area for our hypothetical spectrograph in the null hypothesis (top panel) and including cosmic variance (bottom panel).  A smaller, higher-resolution survey will always improve the chances of a simple detection, but surveys which are too small lose cosmological information because they include fewer modes.}
\label{beamarea}
\end{figure}

\begin{figure}
\centering
\includegraphics[width=\columnwidth]{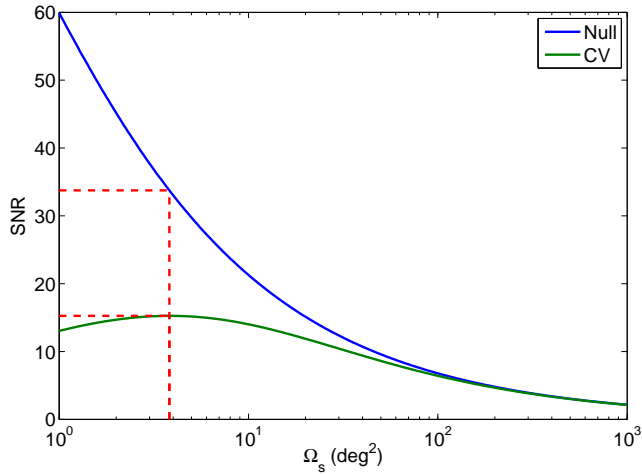}
\caption{SNR as a function of survey area with a beam size of 10 arcmin with and without cosmic variance.  The same behaviors seen in Figure \ref{beamarea} are visible here as well.}
\label{area10}
\end{figure}

Under the null hypothesis, equation (\ref{snrnull}) clearly shows that a smaller, higher resolution survey will always give a higher SNR.  If cosmic variance is included, surveys which are too small will yield smaller SNR's because they include fewer modes.  These behaviors can clearly be seen in Figure \ref{beamarea}, which shows how SNR depends on survey area and beam size, and Figure\ref{area10}, which shows how SNR depends on area for our fiducial 10 arcmin beam.  We choose our fiducial value of the survey area to be $\Omega_s=4$ deg$^2$ because this is the value which maximizes the SNR with cosmic variance included.  This is a much smaller area than what was chosen for a similar spectrograph suggested in P13, and is comparable to the survey area proposed by \citet{vtl}.

\begin{figure}
\centering
\includegraphics[width=\columnwidth]{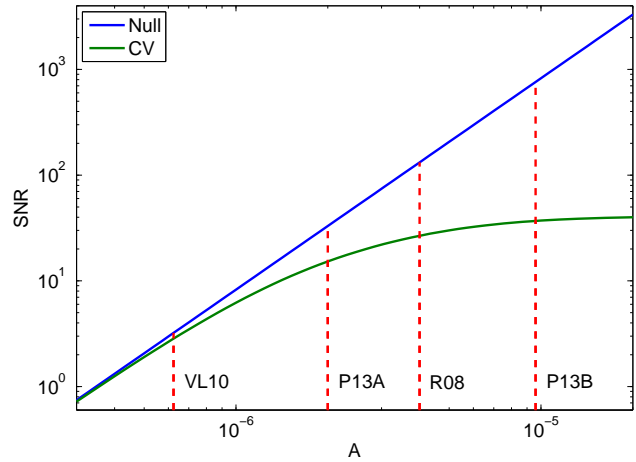}
\caption{signal-to-noise ratio as a function of parameter A for our hypothetical spectrograph with and without cosmic variance.  Values for the four models discussed above are marked.}
\label{SNvA}
\end{figure}

Given the wide variation in the signals predicted by the four models we discussed above, we predict a wide range of possible SNR.  Figure \ref{SNvA} shows the SNR as a function of the parameter $A$, with the values for the four models discussed above marked by dashed red lines.  The curve for the null hypothesis is simply a power law since $SNR\propto A^2$ when cosmic variance is neglected.  Possible values of SNR range from 3.2 (VL10) to 760 (P13B) under the null hypothesis and from 2.8 (VL10) to 37 (P13B) including cosmic variance.  All of the models except the most pessimistic have a good chance to detect the signal.

The predicted SNR is also sensitive to other theoretical parameters in the models such as $\Mmin$ and $t_*$.  These parameters are poorly constrained, so there is some additional uncertainty beyond that shown in Figure \ref{SNvA}.  For example, as seen in Figure 1 of \citet{pullen}, if $\Mmin$ is increased to $10^{10}\ M_{\sun}$ then the average brightness temperature falls off by a factor of $\sim2$.  This would in turn decrease the null hypothesis SNR by a factor of 4.  Changing the value of $t_*$ would change $\fduty$ and have a similar effect on the predicted SNR.

\subsection{Survey Parameters}
\begin{figure}
\centering
\includegraphics[width=\columnwidth]{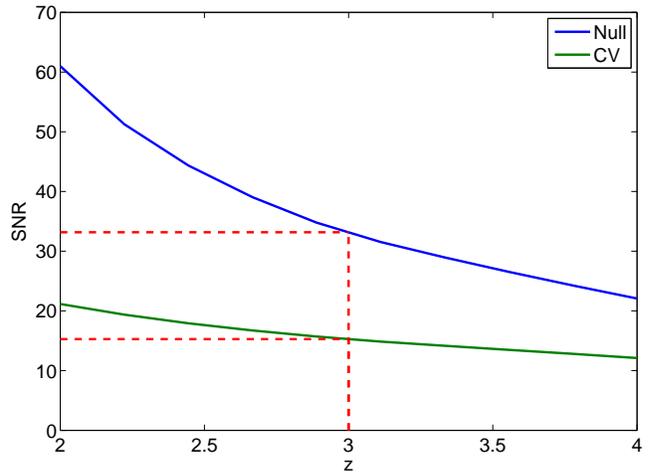}
\caption{signal-to-noise ratio as a function of central redshift for our hypothetical spectrograph with and without the null hypothesis, with the value at $z=3$ marked.  The total bandwidth of the spectrograph is held constant at 1 GHz while the central frequency is varied.}
\label{snvz}
\end{figure}

As shown in Figure \ref{beamarea} above, the parameters of an intensity mapping survey must be carefully chosen to maximize the chance to detect the signal.  Here we explore the dependence of SNR on some of the other survey parameters.  The first possibility we consider is altering the redshift targeted by the survey to see if $z=3$ is actually the best redshift to target.  Figure \ref{snvz} shows the SNR as a function of the central redshift of the survey for our optimal spectrograph.  It is clear that SNR can be increased by targeting lower redshifts.  However, when cosmic variance is included the changes are relatively minor, so the target redshift can be altered somewhat without significantly affecting the SNR.

\begin{figure}
\centering
\includegraphics[width=\columnwidth]{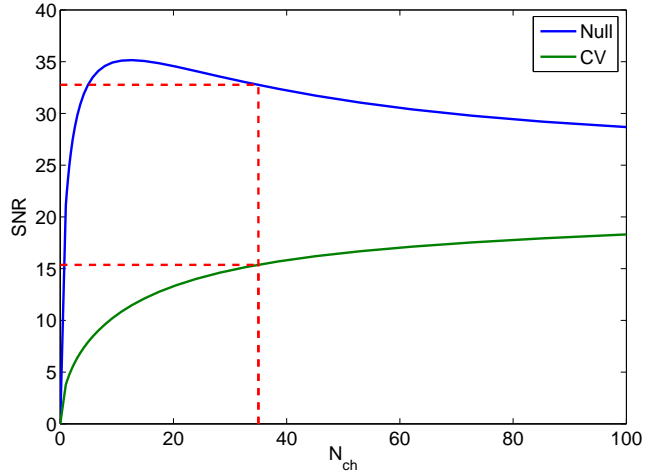}
\caption{Signal to noise as a function of number of spectrograph channels with and without cosmic variance.  The fiducial value $N_{ch}=35$ is marked.  The SNR shown here is obtained by stacking the signals from each of the $N_{ch}$ channels.}
\label{nch}
\end{figure}

Figure \ref{nch} shows the effect of varying the number of frequency channels in the spectrographs.  The total 1 GHz bandwidth is held constant, so if $N_{ch}$ is increased the width of an individual channel is decreased. The shapes of the curves in Figure \ref{nch} are due to several factors.  For small values of $N_{ch}$, increasing the number of channels increases the amplitude of the clustering power spectrum as seen in Figure 1.  As $N_{ch}$ gets larger though, this effect lesses as the clustering power spectrum approaches a constant value.  In addition,  the telescope sensitivity $s$ is proportional to $\Delta\nu^{-1/2}$ \citep{vtl} so the amplitude of the noise power spectrum increases for spectrographs with smaller channels, causing the decrease in SNR seen in the null hypothesis term and the slowed increase seen in the cosmic variance term.  Finally, for very small channels the shot noise power spectrum overtakes the clustering spectrum causing the signal amplitude to increase again, slowing the decrease in SNR in the null hypothesis term.   It can be seen from Figure \ref{nch} that a lower resolution spectrograph produces a higher null hypothesis SNR.  However, this increase is minor and with it comes a decrease in cosmic variance SNR.  Thus it may be preferable to use a spectrograph with $N_{ch}$ near our chosen value.

\begin{figure}
\centering
\includegraphics[width=\columnwidth]{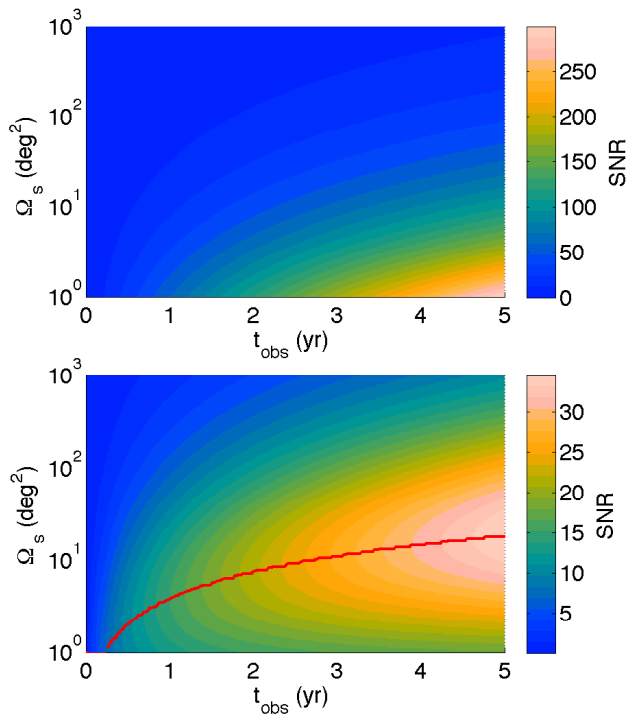}
\caption{Signal to noise as a function of survey time and area for the null hypothesis (top panel) and including cosmic variance (bottom panel).  The red line in the bottom panel shows the optimal survey area for a given observing time assuming a 10 arcminute beam.}
\label{time}
\end{figure}

Another obvious way to improve the SNR of a survey is simply to increase the total observation time.  How much SNR can be gained by observing for longer periods is less obvious.  Longer observing times decrease the amplitude of the noise power spectrum, but eventually the $C_\ell$ cosmic variance term in equation~(\ref{varcv}) starts to dominate over the $C_\ell^n$ term.  At this point, it is more useful to survey a larger area of sky rather than spend additional time on an already deeply studied patch.  This effect is illustrated in figure~\ref{time}, which shows the SNR as a function of survey area and observing time for our two spectrographs, assuming the values calculated above are for a 1 year survey.  For longer observations, the maximum SNR is obtained by surveying a larger area of the sky.

\section{Conclusion}
We have presented a study based on several models of CO emission and a construction of an optimal survey aimed at detecting it.  We briefly discussed four models which estimated the intensity of CO emission using slightly different methods and we found that the large theoretical uncertainties in the calculation lead to a broad range of possible values.  When calculating signal-to-noise ratios for a representative of these models, model P13A, we found that the optimal survey to detect this signal is one which  deeply surveys a relatively small portion of the sky.  We found that the exact target redshift is not too important, and that an instrument with a higher spectral resolution can gain a slight increase in SNR.  The instruments we describe here are able to attain a reasonable SNR given model P13A, and they could provide a much stronger detection considering either model R08 or P13B.  Model VL10 is much more pessimistic, but since it is the most simplistic of the models (relying only on one galaxy for normalization), it appears less likely than the others.

It is important to note that all of our calculations in this paper have taken into account only instrumental noise and cosmic variance.  We have not included any estimates of the impact of foregrounds on the signal-to-noise ratios above.  Since we are looking at line emission, foregrounds with continuous spectra should be fairly easy to remove \citep{vtl}.  However, it is possible for other lines besides the CO line we want to study to be redshifted into the same frequency range.  This line confusion could be mitigated by cross-correlating the CO signal with another map of the same area, either another intensity map in a different frequency or a more traditional map of galaxies or quasars.  Estimating the importance of line confusion is left for future study. 

The results of this work suggest that it is possible to design a survey to detect the CO auto power spectrum if foregrounds are not a major concern.  However, we have shown that the parameters of such a survey should be considered carefully in order to maximize the chance of detection.  Analyses like this will be important if intensity mapping surveys are to reach their full potential.

This work was supported at Johns Hopkins by NSF Grant No. 0244990 and by the John Templeton Foundation.  EDK was supported by the National Science Foundation under Grant Number PHY-1316033.

\end{document}